\documentclass[12pt,a4paper]{amsart}

\usepackage{amssymb}
\makeatletter                                            
\renewcommand{\@begintheorem}[2]{                        
\rm \trivlist \item [\hskip \labelsep {\bf #2\ \ #1.}]   
                                }                        
\makeatother                                             

\makeatletter

\DeclareFontFamily{U}{cyr}{}
\DeclareFontShape{U}{cyr}{m}{n}{
  <5> wncyr5 <6> wncyr6 <7> wncyr7 <8> wncyr8 <9> wncyr9 <10->
wncyr10}{}
\DeclareMathAlphabet{\mathcyr}{U}{cyr}{m}{n}

\input cyracc.def
\input epsf

\newcommand{\ZZ}{{\bf Z}}

\newcommand{\CC}{{\bf C}}

\newcommand{\FF}{{\bf F}}
\newcommand{\HH}{{\bf H}}

\newcommand{\tF}{{\tilde{F}}}

\title{Genus four superstring measures}
\author{Sergio L.~Cacciatori}
\address{Dipartimento di Scienze Fisiche e Matematiche, Universit\`a dell'Insubria,
Via Valleggio 11, I-22100 Como, Italia}
\email{sergio.cacciatori@uninsubria.it}
\author{Francesco Dalla~Piazza}
\address{Dipartimento di Scienze Fisiche e Matematiche, Universit\`a dell'Insubria,
Via Valleggio 11, I-22100 Como, Italia}
\email{f.dallapiazza@uninsubria.it}
\author{Bert van Geemen}
\address{Dipartimento di Matematica, Universit\`a di Milano,
Via Saldini 50, I-20133 Milano, Italia}
\email{geemen@mat.unimi.it}

\begin{document}

\begin{abstract}
A main issue in superstring theory are the superstring measures. D'Hoker and Phong showed that for genus two these reduce to measures on the moduli
space of curves which are determined by modular forms of weight eight and the bosonic measure. They also suggested a generalisation to higher genus.
We showed that their approach works, with a minor modification, in genus three and we announced a positive result also in genus four.
Here we give the modular form in genus four explicitly. 
Recently S. Grushevsky published this result as part of a more general approach.
\end{abstract}

\maketitle

\section{Introduction}

In this letter we present our construction of a candidate for the superstring measure (corresponding to an even spin structure) at four loops.
Following the Ansatz of D'Hoker and Phong in \cite{DP2}, with the modification introduced in \cite{CDG}, this amounts to finding, for each even spin structure, a modular form which satisfies certain transformation as well as certain factorization constraints.
In that paper we found a solution for
the three loop superstring measure and in \cite{DG} its uniqueness is proved. Moreover we announced positive results for the $g=4$ case. Here we explain these results in the $g=4$ case. 
Very recently, while we were still trying to generalise and clarify our $g=4$ results, a paper of S. Grushevsky \cite{Grr} appeared which presents a general proposal for any $g$. However his approach suffers from the fact that it is not obvious at all that his solution is single valued for $g\geq 5$. Recently R.\ Salvati Manni \cite{SM2} obtained results on the single valuedness in case $g=5$.

Our result for the $g=4$ measure is the same as in
\cite{Grr}; however, it has been obtained earlier and independently with a different approach.
Our approach emphasizes the representation of the finite symplectic group on the modular forms (cf.\ \cite{DG}) to pick out the finite dimensional vector space of all modular forms which satisfy the transformation constraints.
Next we try to find a form in this space which also satisfies the factorization constarints.
To actually write down these
forms we use the geometry of isotropic subspaces in order to obtain the correct transformation proprieties of the superstring measure.
Grushevsky, inspired by our first paper, also uses these subspaces.
Our method also leads to uniqueness results.
For these reasons, we decided to present here a shortened version of our proof in the $g=4$ case, following the same approach as in the $g=3$ case in \cite{CDG}. 

The unsurmountable difficulty to prove uniqueness in $g=4$ is the non-normality of the ring of Siegel modular forms in genus four
(see \cite{SM1}). However, we verified uniqueness for four loops in a weakened form in \cite{DG}. 
Grushevsky restricts the search for the $g$ loop amplitudes to a certain vector space of dimension $g+1$ and shows uniqueness in this vector space. However, for genus three (four) the vector space defined by the transformation constraints has dimension five ($\geq 7$ respectively), see \cite{DG} 7.4 and 7.5.

\section{The modular forms $\Xi_8[\Delta]$ for $g\leq 4$}

\subsection{} In \cite{CDG} we suggested,
following an ansatz of D'Hoker and Phong \cite{DP2},
that the genus $g$ superstring measure ${\rm d}\mu[\Delta^{(g)}]$,
for an even characteristic (equivalently, even spin structure)
$$
\Delta^{(g)}=\left[{}^{a_1\ldots a_g}_{b_1\ldots b_g}\right],\qquad a_i,b_i\in\{0,1\},
\qquad \sum a_ib_i\;\equiv\,0\;\mbox{mod}\;2
$$
is the product of the bosonic string measure and a modular form
$\Xi_8[\Delta^{(g)}]$ on Teichm\"uller space of weight $8$.
We observed in \cite{CDG} that the (modified) constraints on $\Xi_8[\Delta^{(g)}]$ imply that $\Xi_8[\Delta^{(g)}]$
is obtained, using the action of the symplectic group 
(which is a quotient of the Teichm\"uller modular group $\Gamma^T_g$), 
from the function $\Xi_8[0]$,
where $[0]=[0^{(g)}]$ is the characteristic with $a_i=b_i=0$ for all $i$.

The function $\Xi_8[0]$ should be invariant
under the action of the subgroup of $\Gamma^T_g$ of elements mapping to
$\Gamma_g(1,2)\subset Sp(2g,\ZZ)$, where $\Gamma_g(1,2)$ is the subgroup
fixing the characteristic $[0]$, under the natural, transitive, action of $Sp(2g,\ZZ)$ on the even characteristics. This implies that
$\Xi_8[0]$ defines a function on the Jacobi locus $J_g\subset\HH_g$ in the Siegel upper half plane  of period matrices of Riemann surfaces, 
recall that the Jacobi locus $J_g$ is the
image of the Teichm\"uller space under the period map.
We will consider $\Xi_8[0]$ as function on $J_g$, actually for $g\;\leq 4$ we will see that it extends naturally to a modular form of weight $8$ on all
of $\HH_g$.
A further condition on $\Xi_8[0]$ is that it factors when the
period matrix becomes reducible:
$$
\Xi_8[0^{(g)}](\tau_{k,g-k})\,=\,
\Xi_8[0^{(k)}](\tau_k)\Xi_8[0^{(g-k)}](\tau_{g-k}),
\qquad
\tau_{k,g-k}\,:=\,
\left(
\begin{array}{cc}
\tau_k & 0 \\
0 & \tau_{g-k}
\end{array}
\right)
\,\in\HH_g,
$$
in which case $\tau_k\in \HH_k,\;\tau_{g-k}\in\HH_{g-k}$.
Finally one requires that in the case $g=1$ one recovers the well-known superstring measure. This implies that in genus one:
$$
\Xi_8[{}^0_0]=\theta[{}^0_0]^4\eta^{12}.
$$
In \cite{CDG} it was shown that these constraints have a solution for $g\leq 3$ and a solution for $g=4$ was also announced, moreover in \cite{DG} we
showed that for $g\leq 3$ the solution is unique. In particular for $g=2$ we do recover the superstring measure as found by D'Hoker and Phong in
\cite{DPg2a}, \cite{DPg2b} using superstring theory (but this can also be verified by direct computation).

\subsection{The case $g=2$}\label{g=2}
In case $g=2$, the following three functions on $\HH_2$ are Siegel modular
forms of weight $6$ on $\Gamma_2(2)$:
$$
f_1:=\theta[{}^{00}_{00}]^{12},\qquad
f_2:=\sum_\delta\theta[\delta]^{12},\qquad
f_3:=\theta[{}^{00}_{00}]^4\sum_\delta\theta[\delta]^{8},
$$
where we sum over the $10$ even characteristics $\delta$ in genus $2$.
The functions $\theta[0^{(2)}]^4f_i$, $i=1,2,3$, are modular forms
of weight $8$ for $\Gamma_g(1,2)$.
In \cite{CDG} we showed that the following function satisfies the constraints
for $g=2$:
$$
\Xi_8[{}^{00}_{00}]\,=\,\theta^4[{}^{00}_{00}](
4f_1+2f_2-3f_3)/6\,=\,\theta^4[{}^{00}_{00}]\Xi_6[{}^{00}_{00}],
$$
with $\Xi_6[0^{(0)}]$ the modular form determined by D'Hoker and Phong
in \cite{DPg2a}, \cite{DPg2b}.
We will show in \cite{DG} that $\Xi_8[0^{(2)}]$ is the unique
modular form on $\Gamma_2(1,2)$ satisfying the constraints.

\subsection{The case $g=3$}\label{g=3}
An obvious generalisation of the functions $f_i$ which we considered
in section \ref{g=2} are:
$$
F_1:=\,\theta[{}^{000}_{000}]^{12},\qquad
F_2:=\,\sum_\Delta \,\theta[\Delta]^{12},\qquad
F_3:=\,\theta[{}^{000}_{000}]^4\sum_\Delta\theta[\Delta]^8,
$$
where the sum is over the $36$ even characteristics $\Delta$ in genus
three. These functions are modular forms of weight $6$ on $\Gamma_3(1,2)$.

Next we use the geometry of the finite symplectic vector space $V=\FF_2^{2g}$ with the standard symplectic form $E$ and the interpretation of even
characteristics with `even'(=split) quadrics, as in \cite{CDG}, to define  modular forms $P_L$ and $G[\Delta]$.

For a Lagrangian  subspace $L$ we define
a modular form of weight $8$ on $\Gamma_3(2)$:
$$
P_L\,:=\,\prod_{Q\supset L}\theta[\Delta_Q]^2,
$$
here the product is over the even quadrics which contain $L$
(there are eight such quadrics for each $L$)
and $\Delta_Q$ is the even characteristic corresponding to $Q$.

For an even characteristic $\Delta$, the quadric $Q_\Delta$ contains
$30$ Lagrangian subspaces.
The sum of the $30$ $P_L$'s, with $L$ a Lagrangian subspace of $Q_\Delta$, is a modular form $G[\Delta]$ of weight $8$ on $\Gamma_3(2)$:
$$
G[\Delta]\,:=\,\sum_{L\subset Q_\Delta}\,P_L\,=\,
\,\sum_{L\subset Q_\Delta}\,
\prod_{Q'\supset L}\theta[\Delta_{Q'}]^2.
$$
Note that $\theta[\Delta]^2$ is one of the factors in each of the 30 products.
In \cite{CDG} we showed that
$$
\Xi_8[{}^{000}_{000}]\,:=\,
\bigl(\theta[{}^{000}_{000}]^4(4F_1+4F_2-3F_3)-12G[{}^{000}_{000}]
\bigr)/12
$$
is a modular form, of weight $8$ on $\Gamma_3(1,2)$,
which satisfies all the constraints. In \cite{DG} we will show
that it is the only modular form of weight $8$ on $\Gamma_3(1,2)$
which satisfies the constraints.


\section{Genus 4}

\subsection{The modular forms $\tF_i$}
An obvious generalisation of the functions $f_i,F_i$ which we considered
earlier in sections \ref{g=2}, \ref{g=3} are:
$$
\tF_1:=\,\theta[{}^{0000}_{0000}]^{12},\qquad
\tF_2:=\,\sum_\Delta \,\theta[\Delta]^{12},\qquad
\tF_3:=\,\theta[{}^{0000}_{0000}]^4\sum_\Delta\theta[\Delta]^8,
$$
where the sum is over the $136$ even characteristics $\Delta$ in genus
four. These functions are modular forms of weight $6$ on $\Gamma_4(1,2)$.
The restriction of these forms to $\HH_1\times\HH_3$ and $\HH_2\times\HH_2$
are easy to find, for example, as
$$
\theta[{}^{abcd}_{efgh}](\tau_{2,2})\,=\,
\theta[{}^{ab}_{ef}](\tau_{2})\theta[{}^{cd}_{gh}](\tau'_{2})
$$
we get:
{\renewcommand{\arraystretch}{1.5}
$$
\begin{array}{rcl}
\tF_1(\tau_{2,2})&=&
\theta[{}^{00}_{00}]^{12}(\tau_2)\theta[{}^{00}_{00}]^{12}(\tau'_2)\\
&=&f_1(\tau_2)f_1(\tau'_2),\\
\tF_2(\tau_{2,2})&=&
\Bigl(\sum_\delta \,\theta[\delta]^{12}\Bigr)(\tau_2)
\Bigl(\sum_\delta \,\theta[\delta]^{12}\Bigr)(\tau'_2)\\
&=&f_2(\tau_2)f_2(\tau'_2),\\
\tF_3(\tau_{2,2})&=&
\theta[{}^{00}_{00}]^{4}(\tau_2)\theta[{}^{00}_{00}]^{4}(\tau'_2)
\Bigl(\sum_\delta \,\theta[\delta]^{8}\Bigr)(\tau_2)
\Bigl(\sum_\delta \,\theta[\delta]^{8}\Bigr)(\tau'_2)\\
&=&f_3(\tau_2)f_3(\tau'_2).
\end{array}
$$
} 

\subsection{The modular form $G_1[0]$}\label{g1to22}
The modular form $G$ can be generalised in two ways. The first is to stick to three dimensional isotropic subspaces $W$ in $\FF_2^8$. Given such a $W$,
there are $3\cdot 8=24$ even quadrics $Q_\Delta$ such that
$W\subset Q_\Delta$. Let $Q_0\subset \FF_2^8$ be the even quadric
with characteristic $\Delta_0=[0^{(4)}]$.
We will only use the octets of quadrics which
contain $Q_0$ to define a modular form $G_1[0]$:
$$
G_1[{}^{0000}_{0000}]\,:=\,\sum_{W\subset Q_0}\,\,\prod_{w\in W}\,
\theta[\Delta_0+w]^2,
$$
where we sum over the $2025$ three dimensional isotropic subspaces
$W\subset Q_0$, 
and for each such subspace we take the product of
the eight even $\theta[\Delta_0+w]^2$.
As $0\in W$, for any subspace $W$, 
the function $G_1[0]$ is a multiple of
$\theta[\Delta_0]^2$. 
The function $G_1[0]$ 
is a modular form on $\Gamma_4(1,2)$ of weight $8$ (cf.\ \cite{Igusatrans}
or use the explicit transformation theory of theta functions as in the Appendices of \cite{CDG} or see \cite{Grr}, Proposition 13).

Using methods similar to those in Appendix C of \cite{CDG} one finds
the restriction of $G_1[0]$ to $\HH_1\times\HH_3$:
{\renewcommand{\arraystretch}{1.5}
$$
\begin{array}{rl}
G_1[{}^{0000}_{0000}](\tau_{1,3})=&
\theta[{}^0_0]^{16}(\tau_1)G[{}^{000}_{000}](\tau_3)+
\Bigl(\theta[{}^0_0]^8(\theta[{}^0_1]^8+\theta[{}^1_0]^8)\Bigr)
(\tau_1)
\Bigl(H[{}^{000}_{000}]+7G[{}^{000}_{000}]\Bigr)(\tau_3)\\
=&
\theta[{}^0_0]^{4}(\tau_1)\Bigl((
\mbox{$\frac{1}{3}$}f_{21}+\eta^{12})(\tau_1)
G[{}^{000}_{000}](\tau_3)+
(\mbox{$\frac{1}{3}$}f_{21}-\eta^{12})
(\tau_1)
(H[{}^{000}_{000}]+7G[{}^{000}_{000}](\tau_3)\Bigr)\\
=&
\theta[{}^0_0]^4(\tau_1)
\Bigl(\mbox{$\frac{1}{3}$}f_{21}(\tau_1)
(H[{}^{000}_{000}]+8G[{}^{000}_{000}])(\tau_3)
-\eta^{12}(\tau_1)(H[{}^{000}_{000}]+6G[{}^{000}_{000}])(\tau_3)\Bigr)
\end{array}
$$
} 
where 
$$
f_{21}\,:=\,2\theta[{}^0_0]^{12}+\theta[{}^0_1]^{12}+\theta[{}^1_0]^{12}, \qquad
H[{}^{000}_{000}]\,:=\,\sum_{W'\subset Q_0}
\prod_{w\in W}\,
\theta[\Delta_0^{(3)}+w]^4,
$$
for $f_{21}$ see \cite{CDG}, section 3.3, the sum in $H$ is over the $105$ isotropic $2$-dimensional subspaces
$W'$ contained in $Q_0$, where now $Q_0\subset\FF_2^3$.

Similarly,
the restriction of $G_1[0]$ to $\HH_2\times\HH_2$ is
$$
G_1[{}^{0000}_{0000}](\tau_{2,2})\,=\,
\theta[{}^{00}_{00}]^4(\tau_2)\theta[{}^{00}_{00}]^4(\tau'_2)
\Bigl(
g(\tau_2)f_3(\tau_2')+f_3(\tau_2)g(\tau'_2)+
9g(\tau_2)g(\tau'_2)
\Bigr),
$$
with $f_3$ as in section \ref{g=2} and:
$$
g(\tau_2)\,=\,
\sum_{W'\subset Q_0}\,\prod_{w\in W'-\{0\}}\theta[\Delta_0^{(2)}+w]^4(\tau_2),
$$
and the sum in $g$ is over the $6$ isotropic 2-dimensional subspaces
$W'$ of $Q_0$, where now $Q_0\subset\FF_2^4$, and we take a product of only three terms (the factor $\theta[0^{(2)}]^4$ for $w=0$ is taken out in the formula
for $G_1[0](\tau_{2,2})$).

\subsection{The modular form $G_2[\Delta]$}
Another generalisation of $G$ uses Lagrangian subspaces $L\cong\FF_2^4$ of $V=\FF_2^8$. For each $L$ there are $16$ even quadrics $Q_\Delta$ with
$L\subset Q_\Delta$. For an even characteristic $\Delta$ we define
$$
G_2[\Delta]\,=\,\sum_{L\subset Q_\Delta}\,\prod_{Q\supset L}
\theta[\Delta_Q],
$$
here we take the sum over the $270$ (four dimensional) Lagrangian subspaces $L$ of $V=\FF_2^8$ which are contained in $Q_\Delta$.
Again, $L\subset Q_\Delta$ implies that $G_2[\Delta]$ is a multiple of
$\theta[\Delta]$.
The function $G_2[0]$ 
is also modular form on $\Gamma_4(1,2)$ of weight $8$.

The restriction of $G_2[0]$ to $\HH_1\times\HH_3$
is:
{\renewcommand{\arraystretch}{1.5}
$$
\begin{array}{rcl}
G_2[{}^{0000}_{0000}](\tau_{1,3})&=&
\Bigl(\theta[{}^0_0]^8
(\theta[{}^0_1]^8+\theta[{}^1_0]^8)\Bigr)(\tau_1)
G[{}^{000}_{000}](\tau_3)\\
&=&\Bigl(\theta[{}^0_0]^4(\mbox{$\frac{1}{3}$}f_{21}-\eta^{12})\Bigr)
(\tau_1)G[{}^{000}_{000}](\tau_3).
\end{array}
$$
}  
The restriction of $G_2[0]$ to $\HH_2\times\HH_2$ is as follows:
$$
G_2[{}^{0000}_{0000}](\tau_{2,2})\,=\,
\theta[{}^{00}_{00}]^4(\tau_2)\theta[{}^{00}_{00}]^4(\tau'_2)
g(\tau_2)g(\tau'_2),
$$
with $g$ as in \ref{g1to22}.

\subsection{Lower genus identities}\label{g3id}
Before considering the restriction to $\HH_1\times\HH_3$ we give some identities which we
need to give the restriction of the $G_i[0]$'s as a linear combination of the $F_i$'s and $G[\Delta]$ in genus three (see \ref{g=3}) and the $f_i$'s in genus two (see \ref{g=2}). For $H[0^{(3)}]$ and $g$ (see \ref{g1to22}) we have:
$$
H[{}^{000}_{000}]\,=\,\theta[{}^{000}_{000}]^{4}(2F_1+
8F_2 -3F_3)/6, \qquad g\,=\,(2f_1+4f_2-3f_3)/6.
$$
These identities can be verified using the classical theta formula
stated in \cite{CDG}, section 3.2 (it is helpful to use a computer as well).

\subsection{Constraints for $\Xi_8[\Delta]$}
We now want to find
a modular form $\Xi_8[0^{(4)}]$ of weight $8$ on
$\Gamma_4(1,2)$
which restricts to $\HH_1\times\HH_3$ as
$$
\Xi_8[{}^{0000}_{0000}](\tau_{1,3})\,=\,
\Xi_8[{}^0_0](\tau_1)\Xi_8[{}^{000}_{000}](\tau_3)\,=\,
\Bigl(\theta[{}^0_0]^4\eta^{12}\Bigr)(\tau_1)
\Xi_8[{}^{000}_{000}](\tau_3)
$$
and which restricts to $\HH_2\times\HH_2$ as
$$
\Xi_8[{}^{0000}_{0000}](\tau_{2,2})\,=\,
\Xi_8[{}^{00}_{00}](\tau_2)\Xi_8[{}^{00}_{00}](\tau_2')\,=\,
\Bigl(\theta[{}^{00}_{00}]^4\Xi_6[{}^{00}_{00}]\Bigr)(\tau_2)
\Bigl(\theta[{}^{00}_{00}]^4\Xi_6[{}^{00}_{00}]\Bigr)(\tau_2').
$$

\subsection{The strategy.}
We consider a general linear combination of the $5$ functions $\theta[0^{(4)}]^4\tF_1$, $\theta[0^{(4)}]^4\tF_2$, $\theta[0^{(4)}]^4\tF_3$, $G_1[0]$, $G_2[0]$ and try to find one whose restriction to
$\HH_1\times\HH_3$ is $\Xi_8[0^{(1)}]\Xi_8[0^{(3)}]$.
For this, we use that, as function of $\tau_1\in\HH_1$, this restriction
is a linear combination of $\Xi_8[0^{(1)}]$ and $\theta[0^{(1)}]^4f_{21}$
so we require that  the last term should vanish.
It then happens that what remains is exactly
$\Xi_8[0^{(1)}]\Xi_8[0^{(3)}]$. 
Finally we verified that
the restriction of this function to $\HH_2\times\HH_2$
is indeed $\Xi_8[0^{(2)}](\tau_2)\Xi_8[0^{(2)}](\tau'_2)$.

\subsection{The restriction to $\HH_1\times\HH_3$}
We have the following formulas for the restriction:
{\renewcommand{\arraystretch}{1.5}
$$
\begin{array}{rl}
(\theta[{}^{0000}_{0000}]^{4}\tF_1)_{|_{\HH_1\times\HH_3}}=&
\theta[{}^0_0]^4
\Bigl(
(\mbox{$\frac{1}{3}$}f_{21}+\eta^{12})\theta[{}^{000}_{000}]^{4}F_1
\Bigr),
\\&\\
(\theta[{}^{0000}_{0000}]^4\tF_2)_{|_{\HH_1\times\HH_3}}=&
\theta[{}^0_0]^4
\Bigl(
(\mbox{$\frac{2}{3}$}f_{21}-\eta^{12})
\theta[{}^{000}_{000}]^4 F_2\Bigr),
\\&\\
(\theta[{}^{0000}_{0000}]^4\tF_3)_{|_{\HH_1\times\HH_3}}
=&
\theta[{}^0_0]^4
\Bigl(
\mbox{$\frac{2}{3}$}f_{21}
\theta[{}^{000}_{000}]^4 F_3
\Bigr),
\\&\\
(G_2[{}^{0000}_{0000}])_{|_{\HH_1\times\HH_3}}=&
\theta[{}^0_0]^4\Bigl((\mbox{$\frac{1}{3}$}f_{21}-\eta^{12})
G[{}^{000}_{000}]\Bigr),
\\&\\
(G_1[{}^{0000}_{0000}])_{|_{\HH_1\times\HH_3}}=&
\theta[{}^0_0]^4
\Bigl(\mbox{$\frac{1}{3}$}f_{21}
(H[{}^{000}_{000}]+8G[{}^{000}_{000}])
-\eta^{12}(H[{}^{000}_{000}]+6G[{}^{000}_{000}])\Bigr).
\end{array}
$$
} 

\subsection{The constraint on $\HH_1\times\HH_3$}\label{eqng3}
A linear combination $a_1\theta[0^{(4)}]^{4}\tF_1+\ldots+a_4G_2[0]+a_5G_1[0]$
of the five functions above
is a multiple of $\eta^{12}(\tau_1)$ iff
$$
a_1\theta[{}^{000}_{000}]^{4}F_1+
2a_2\theta[{}^{000}_{000}]^4F_2 +
2a_3\theta[{}^{000}_{000}]^4F_3 +
a_4G[{}^{000}_{000}]+
a_5(H[{}^{000}_{000}]+8G[{}^{000}_{000}])=0.
$$
Using the formula for $H$ given in \ref{g3id}
and denoting $b_5:=a_5/6$ to simplify formulas, we get:
$$
(a_1+2b_5)\theta[{}^{000}_{000}]^{4}F_1+
(2a_2+8b_5)\theta[{}^{000}_{000}]^4F_2+
(2a_3-3b_5)\theta[{}^{000}_{000}]^4F_3 +
(a_4+48b_5)G[{}^{000}_{000}]=0.
$$
As the four functions here are independent (cf.\ \cite{DG}), we get the solutions:
$$
(a_1,a_2,a_3,a_4,a_5)\,=\,\lambda(-2,-4,3/2,-48,6),\qquad
(\lambda\in\CC).
$$
For such $a_i$ the linear combination $a_1\theta[0^{(4)}]^{4}\tF_1+\ldots+a_4G_2[0]+a_5G_1[0]$ restricts to:
$$
\theta[{}^0_0]^4\eta^{12}\Bigr(a_1\theta[{}^{000}_{000}]^{4}F_1
-a_2\theta[{}^{000}_{000}]^4F_2
-a_4G[{}^{000}_{000}]
-a_5(H[{}^{000}_{000}]+6G[{}^{000}_{000}])\Bigr)
$$
which, using again the formula for $H$, and $a_5=6b_5$,
gives a genus three factor given by:
{\renewcommand{\arraystretch}{1.5}
$$
(a_1-2b_5)\theta[{}^{000}_{000}]^{4}F_1
-(a_2+8b_5)\theta[{}^{000}_{000}]^4F_2+
3b_5\theta[{}^{000}_{000}]^4F_3
-(a_4+36b_5)G[{}^{000}_{000}].
$$
} 
Putting
$\lambda=-2$, so $(a_1,\ldots,a_5)=(4,8,-3,96,-12)$ and $b_5=-2$, we get
(cf.\ \ref{g=3})
$$
8\theta[{}^{000}_{000}]^{4}F_1
+8\theta[{}^{000}_{000}]^4F_2
-6\theta[{}^{000}_{000}]^4F_3
-24G[{}^{000}_{000}]\,=\,24\Xi_8[{}^{000}_{000}].
$$
Thus we found that the function
$$
\Xi_8[{}^{0000}_{0000}]\,:=\,
\Bigl(\theta[{}^{0000}_{0000}]^4(4\tF_1+8\tF_2-3\tF_3)
+96G_2[{}^{0000}_{0000}]-12G_1[{}^{0000}_{0000}]\Bigr)/24
$$
has the property that it satisfies the constraint on the restriction
to $\HH_1\times\HH_3$.

\subsection{The constraint on $\HH_2\times\HH_2$}\label{eqn22}
Now we consider the restriction of the function
$\Xi_8[0^{(4)}]$ from section \ref{eqng3} to $\HH_2\times\HH_2$. It is useful to observe that:
$$
\begin{array}{rcl}
24\Xi_8[{}^{0000}_{0000}](\tau_{2,2})&=&
\theta[{}^{00}_{00}]^4(\tau_2)\theta[{}^{00}_{00}]^4(\tau_2')
h(\tau_2,\tau'_2),
\end{array}
$$
with a holomorphic function $h$ which we will not write down explicitly.
Taking out the factor $\theta[0^{(2)}]^4(\tau_2)\theta[0^{(2)}]^4(\tau_2')$, which also
occurs in $\Xi_8[0^{(2)}](\tau_2)\Xi_8[0^{(2)}](\tau_2')$, simplifies the computation. We found, using the classical theta formula (cf.\ \cite{CDG}, section 3.2) and  a computer, that $h(\tau_2,\tau'_2)=\Xi_6[0^{(2)}](\tau_2)\Xi_6[0^{(2)}](\tau_2')$ and thus:
$$
\Xi_8[{}^{0000}_{0000}](\tau_{2,2})\,=\,
\Xi_8[{}^{00}_{00}](\tau_2)\Xi_8[{}^{00}_{00}](\tau_2').
$$
Therefore the modular form $\Xi_8[0^{(4)}]$ on $\Gamma_4(1,2)$ of weight 8, defined in \ref{eqng3}, satisfies
all the factorization constraints in genus four. 



\end{document}